\documentclass[aps,prc,twocolumn,groupedaddress,showpacs]{revtex4}

\usepackage{graphicx}
\usepackage{dcolumn}
\usepackage{bm}

\begin{document}


\title{Comparison of hadronic rescattering calculations of
elliptic flow and HBT with measurements from RHIC}


\author{T. J. Humanic}
\email[]{humanic@mps.ohio-state.edu}
\homepage[]{http://vdgus1.mps.ohio-state.edu/}
\affiliation{Department of Physics, The Ohio State University,
Columbus, OH 43210}


\date{\today}

\begin{abstract}
Results from the data obtained in the first physics run of the
Relativistic Heavy Ion Collider (RHIC)
have shown suprisingly large elliptic flow and suprisingly small
HBT radii. Attempts to explain both results in a consistant
picture have so far been unsuccessful. The present work shows 
that a simple thermal-like initial state model coupled to a 
hadronic rescattering calculation can explain reasonably well
both elliptic flow and HBT 
results from RHIC. The calculation 
suggests a very early
hadronization time of about 1 fm/c after the initial
collision of the nuclei.
\end{abstract}

\pacs{25.75.Gz, 25.75.Ld}

\maketitle


\section{Introduction}

Results of the Year-1 running of the Relativistic Heavy Ion Collider
(RHIC) for Au+Au collisions at $\sqrt{s}=130$ GeV have shown 
suprisingly large pion elliptic flow \cite{Adler:2001a} and 
suprisingly small
radii from two-pion Hanbury-Brown-Twiss interferometry (HBT)
\cite{Adler:2001b,Adcox:2002a}. 
Attempts to explain both results 
in a consistant picture have so far been unsuccessful. Hydrodynamical
models agree with the large elliptic flow seen in the RHIC data 
\cite{Kolb:1999a} but significantly disagree with the experimental HBT
radii \cite{Rischke:1996a}. On the other hand, relativistic quantum
molecular dynamics calculations which include hadronic 
rescattering, for example RQMD v2.4 \cite{Sorge:1989a},
significantly underpredict the elliptic flow seen in the RHIC data
\cite{Sorge:1995a} but predict pion HBT radii comparable to the data 
\cite{Hardtke:1999a}.
A calculation has recently been made to extract HBT radii 
with a hydrodynamical
model coupled with a hadronic rescattering afterburner with the result
that the HBT radii are significanly larger than measurements
\cite{Soff:2000a}. This lack of a single model to explain both results
has been our first big mystery from RHIC. It has been
suggested that we should call into question 
our current understanding of what information 
pion HBT measurements give us \cite{Gyulassy:2001a}.

In an effort to address this mystery, the present work explores
a somewhat different picture of the nuclear collision than those
presented above. In this picture, hadronization into a simple 
thermal-like
state occurs soon (about 1 fm/c) after the initial collision of the
nuclei followed by hadronic rescattering until freezeout. The goal
will thus be to test whether hadronic rescattering alone
can generate enough general flow in the system to explain both 
the elliptic flow and HBT results from RHIC starting from a
simple non-flowing initial state. Note that this approach,
while similar to the relativistic quantum molecular dynamics
calculations mentioned above, differs from them in the choice of 
the model of the initial state of the system before hadronic 
rescattering commences (e.g. the other models use a color string
picture for the initial state) \cite{Sorge:1989a}.
The simple initial state model used here is meant to be a construction
to parameterize the true initial state with a few parameters
which can be adjusted to help the rescattering calculation agree with
the RHIC results. If it is possible to obtain reasonable agreement
with the data, then one might ask how early in the calculation
one can go and still maintain a physically motivated model. 
The discussion of
this important point is deferred until later.

\section{Calculational Method}

A brief description of the rescattering model
calculational method is given
below. The method used is similar to that used in previous 
calculations for lower CERN Super Proton Synchrotron (SPS)
energies \cite{Humanic:1998a}. 
Rescattering is simulated with a semi-classical 
Monte Carlo calculation which assumes strong binary collisions 
between hadrons. The Monte Carlo calculation is
carried out in three stages: 1) initialization and hadronization, 2)
rescattering and freeze out, and 3) calculation of experimental 
observables. Relativistic kinematics is used 
throughout.  All calculations are made to simulate RHIC-energy
Au+Au collisions in order to compare with the results of the
Year-1 RHIC data.

The hadronization model employs simple parameterizations to describe the 
initial momenta and space-time of the hadrons similar to
that used by Herrmann and Bertsch \cite{Herrmann:1995a}. The initial 
momenta are assumed to follow a thermal transverse
(perpendicular to the beam direction)
momentum distribution for all particles,
\begin{equation}
(1/{m_T})dN/d{m_T}=C{m_T}/[\exp{({m_T}/T)} \pm 1]
\end{equation}
where ${m_T}=\sqrt{{p_T}^2 + {m_0}^2}$ is the transverse mass, $p_T$ 
is the transverse momentum, $m_0$ is the particle rest mass, $C$ is 
a normalization constant, and $T$ is the initial ``temperature''
of the system, 
and a gaussian rapidity distribution for mesons,
\begin{equation}
dN/dy=D \exp{[-{(y-y_0)}^2/(2{\sigma_y}^2)]}
\end{equation}
where $y=0.5\ln{[(E+p_z)/(E-p_z)]}$ is the rapidity, $E$ is the 
particle energy, 
$p_z$ is the longitudinal (along the beam direction)
momentum, $D$ is a normalization constant, 
$y_0$ is the central
rapidity value (mid-rapidity), and $\sigma_y$ is the rapidity width.
Two rapidity distributions for baryons have been tried: 1) flat
and then falling off near beam rapidity and 2) peaked at central
rapidity and falling off until beam rapidity. Both baryon
distributions give about the same results. 
The initial space-time of the
hadrons for $b=0$ fm (i.e. zero impact parameter or central collisions) 
is parameterized as having cylindrical symmetry with 
respect to the 
beam axis. The transverse particle density dependence is assumed 
to be that of a
projected uniform sphere of radius equal to the projectile radius, $R$ 
($R={r_0}A^{1/3}$, where ${r_0}=1.12$ fm and $A$ is the
atomic mass number 
of the projectile). For $b>0$ (non-central collisions) the transverse
particle density is that of overlapping projected spheres whose
centers are separated by a distance b.
The longitudinal
particle hadronization position ($z_{had}$) and time ($t_{had}$) 
are determined by the relativistic equations \cite{Bjorken:1983a},
\begin{equation}
z_{had}=\tau_{had}\sinh{y};   t_{had}=\tau_{had}\cosh{y}
\end{equation}
where $y$ is the particle rapidity and $\tau_{had}$ is the 
hadronization proper time.
Thus, apart from particle multiplicities, the hadronization model has 
three free
parameters to extract from experiment: $\sigma_y$,
$T$ and $\tau_{had}$.
The hadrons included in the calculation are pions, kaons,
nucleons and lambdas
($\pi$, K, N, and $\Lambda$), and the $\rho$, $\omega$, $\eta$, 
${\eta}'$, 
$\phi$, $\Delta$, and $K^*$ resonances. For simplicity, the
calculation is isospin averaged (e.g. no distinction is made among
a $\pi^{+}$, $\pi^0$, and $\pi^{-}$). Resonances are present at 
hadronization
and also can be produced as a result of rescattering. Initial resonance
multiplicity fractions are taken from 
Herrmann and Bertsch \cite{Herrmann:1995a}, 
who extracted results from the HELIOS experiment \cite{Goerlach:1992a}. 
The initial resonance fractions used in
the present calculations are: $\eta/\pi=0.05$, $\rho/\pi=0.1$, 
$\rho/\omega=3$, $\phi/(\rho+\omega)=0.12$,
${\eta}'/\eta=K^*/\omega=1$ and, for simplicity, $\Delta/N=0$.

The second stage in the calculation is rescattering 
which finishes with the
freeze out and decay of all particles. Starting 
from the initial stage ($t=0$ fm/c), the positions 
of all particles are allowed to evolve in time in small
time steps ($dt=0.1$ fm/c) according to their 
initial momenta. At each time step
each particle is checked to see a) if it decays, and b) if it is 
sufficiently
close to another particle to scatter with it.
Isospin-averaged s-wave and p-wave cross sections 
for meson scattering are 
obtained from Prakash et al.
\cite{Prakash:1993a}. The calculation is carried out to 100 fm/c,
although most of the rescattering finishes by about 30 fm/c.
The rescattering calculation
is described in more detail elsewhere \cite{Humanic:1998a}.

Calculations are carried out assuming initial parameter values and particle
multiplicities for each type of particle. In the last stage of the 
calculation, the freeze-out and decay momenta and 
space-times are used to produce
observables such as pion, kaon, and nucleon
multiplicities and transverse momentum and rapidity distributions. 
The values of the 
initial parameters of the calculation and multiplicities
are constrained to give observables which agree with 
available measured hadronic observables. As a cross-check
on this, the total kinetic energy from the calculation is 
determined and
compared with the RHIC center of mass energy of 
$\sqrt{s}=130$ GeV to see that they are in 
reasonable agreement. Particle multiplicities were estimated from
the charged hadron multiplicity measurements of the RHIC 
PHOBOS experiment \cite{Back:2000a}. Calculations were carried
out using isospin-summed events containing at
freezeout about 5000 pions, 500 kaons, and 650 nucleons
($\Lambda$'s were decayed).
The hadronization model parameters used were $T=300$ MeV,
$\sigma_y$=2.4, and $\tau_{had}$=1 fm/c. It is interesing
to note that the same value of $\tau_{had}$ was required 
in a previous rescattering calculation to successfully 
describe results from SPS
Pb+Pb collisions \cite{Humanic:1998a}. 

Figure 1 shows normalized pseudorapidity distributions
summed over pions, kaons, and nucleons from
rescattering calculations for $b=0, 5$, and 8 fm. The widths
of the distributions and the flattening near $\eta=0$ is similar
to data from PHOBOS \cite{Back:2001a}. Note that if the initial
pion rapidity distribution in the calculation was slightly 
flattened at $y=0$ from the simple Gaussian in Equation 2, the
small dip seen in the data at $\eta=0$ could be reproduced.
\begin{figure}[t]
\includegraphics{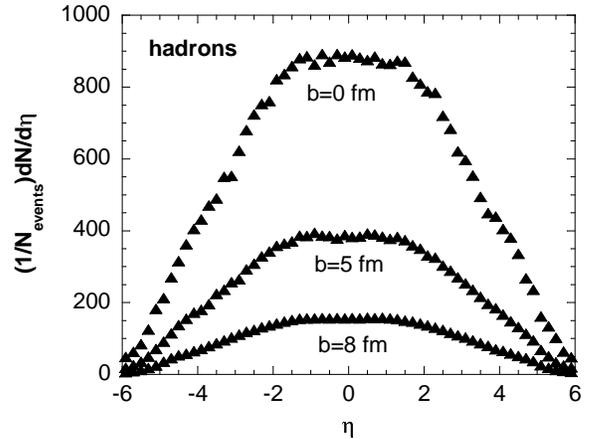}
\caption{\label{fig:pseudo} Pseudorapidity distributions summed
over pions, kaons, and nucleons from rescattering calculations
for $b=0, 5$, and 8 fm.}
\end{figure}

Figure 2 shows
$m_T$ distributions for pions, kaons, and nucleons from the
rescattering calculation for $b=0$ fm near midrapidity ($-1<y<1$)
fitted to exponentials
of the form $\exp{(-m_T/B)}$, where $B$ is the slope parameter.
\begin{figure}
\includegraphics{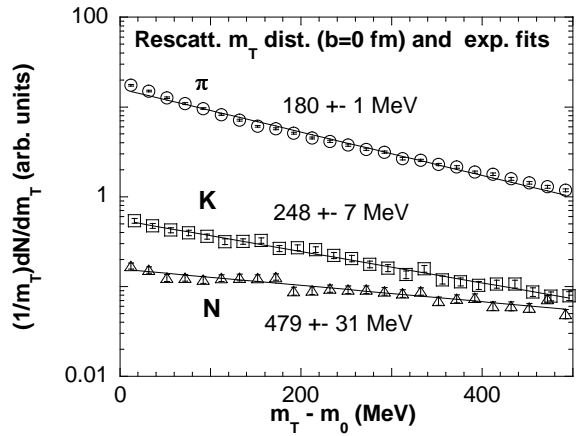}
\caption{\label{fig:mT} Transverse mass distributions from the
rescattering model. The lines are exponential fits to the
distributions and the slope parameters are shown.}
\end{figure}
The extracted slope parameters shown in Figure 2 are close in value
to preliminary measurements from the STAR experiment for the
$\pi^-$, $K^-$, and anti-proton of $190\pm10$, 
$300\pm30$,
and $565\pm50$ MeV, respectively \cite{Adler:2002a}. Thus, we see that
if all hadrons begin at a common temperature-parameter value of 300 MeV,
the hadronic rescattering alone is able to generate enough radial
flow to account for the differences in slope among the 
pion, kaon, and nucleon $m_T$ distributions.

Figure 3 shows transverse momentum 
distributions for pions, kaons,
and nucleons from a $b=8$ fm rescattering calculation which 
extend to high-$p_T$, i.e. 6 GeV/c. As observed
in PHENIX data \cite{Zajc:2001a}, the pion and nucleon distributions merge
for $p_T>2$ GeV/c and the kaon distribution crosses that for nucleons
at around 1 GeV/c.

The elliptic flow 
and two-pion HBT observables
are also calculated from the freeze-out
momenta and space-time positions 
of the particles at the end of the rescattering stage.
The elliptic flow variable, $v_2$, is defined 
as \cite{Poskanzer:1998a}
\begin{equation}
v_2 = \langle cos(2\phi) \rangle;   \phi = \arctan(p_y/p_x)
\end{equation}
where $p_x$ and $p_y$ are the $x$ and $y$ components of the particle
momentum, and $x$ is in the impact parameter direction and
$y$ is in the ``out of plane'' direction (i.e. $x-z$ is the
reaction plane and $z$ is the beam direction).
The HBT pion source parameters are extracted from the
rescattering calculation using the same method as was applied
for previous SPS-energy rescattering calculations \cite{Humanic:1998a}.
The Pratt-Bertsch ``out-side-long'' radius parameterization is used
\cite{Pratt:1990a,Bertsch:1989a} yielding the four parameters
$R_{Tside}$, $R_{Tout}$, $R_{Long}$, and $\lambda$, which
represent two mutually perpendicular transverse 
(to the beam direction) radius parameters,
a radius parameter along the beam direction, and
a parameter related to the ``strength'' of the two-pion 
correlations, respectively.
\begin{figure}
\includegraphics{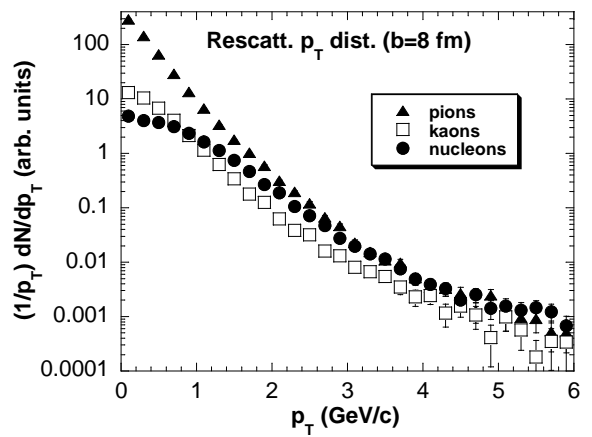}
\caption{\label{fig:pT} Transverse momentum distributions for pions,
kaons, and nucleons from a $b=8$ fm rescattering calculation.}
\end{figure}

\section{Elliptic Flow results}

Figure 4 shows the $p_T$ dependence of $v_2$ for pions and
nucleons extracted from the $b=8$ fm rescattering calculation
compared with the trends of the STAR 
measurements for $\pi^{+}+\pi^{-}$ and $p$ + $p-bar$ 
at $11-45\%$ centrality \cite{Adler:2001a}, which 
roughly corresponds to this impact parameter.
Figure 5 compares the $p_T$ dependence of $v_2$ for kaons from the
$b=8$ fm rescattering calculation with the STAR measurements
for ${K^0}_s$ at $11-45\%$ centrality \cite{Adler:2002b}.
As seen, the 
rescattering calculation values
are in reasonable agreement with the STAR measurements.
The flattening out of the pion and nucleon $v_2$ distributions 
for $p_T>2$ GeV/c is consistant with that seen in 
STAR and PHENIX results for
minimum-bias hadrons \cite{Snellings:2001a,Zajc:2001a} (the kaon
$v_2$ calculation does not extend higher than 2 GeV/c in Figure 5
due to limited statistics).
Thus, the same
rescattering mechanism that can account for the radial flow seen
in Figures 2 and 3 is also seen to account for the magnitude 
and $p_T$ dependence
of the elliptic flow for pions, kaons, and nucleons.

\section{HBT results}

The pion source parameters extracted from HBT analyses of
rescattering calculations for three different impact
parameters, $b=0$, $5$, and $8$ fm, are compared with 
STAR $\pi^{-}$ measurements at three
centrality bins \cite{Adler:2001b} in Figure 6.
Note that the PHENIX HBT results \cite{Adcox:2002a} 
are in basic agreement with the STAR results. 
The STAR
centrality bins labeled ``3'', ``2'', and ``1'' in the figure
correspond to $12\%$ of central, the next $20\%$, and the 
next $40\%$,
respectively. These bins are roughly approximated by the impact
parameters used in the rescattering calculations,
i.e. the average impact parameters of the STAR centrality
bins are estimated to be within $\pm2$ fm of the
rescattering calculation impact parameters used to
compare with them.
In the left
panel, the centrality dependence of the HBT parameters
is plotted for a $p_T$ bin of $0.125-0.225$ GeV/c. In the right
panel, the $m_T$ dependence of the HBT parameters is plotted
for centrality bin 3, for the STAR measurments, or $b=0$ fm,
for the rescattering calculations. Although there are
differences in some of the details, the trends of the STAR
HBT measurements are seen to be described rather well by the
rescattering calculation.
\begin{figure}
\includegraphics{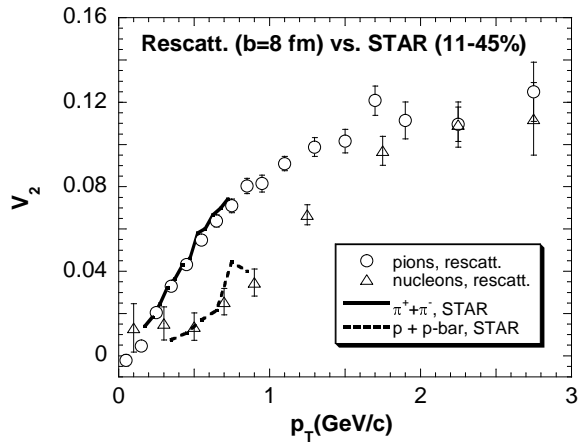}
\caption{\label{fig:v2} Calculations of $v_2$ from the
rescattering model for $b=8$ fm for pions and nucleons
compared with STAR measurements at 11-45\% centrality. 
The plotted points with error 
bars are the
rescattering calculations and the lines show the trends of the
STAR measurements. Average errors on the STAR measurements
are $\leq0.002$ for pions and $0.006$ for protons+antiprotons.}
\end{figure}

\section{Discussion}

As shown above, the radial and elliptic flow as well 
as the features of the HBT 
measurements at RHIC can be adequately described by
the rescattering model with the hadronization model parameters
given earlier. The results of the calculations are found
to be sensitive to the value of $\tau_{had}$ used, as was
studied in detail for SPS rescattering calculations 
\cite{Humanic:1998a}. For calculations with $\tau_{had}>1$ fm/c
the initial hadron density is smaller, fewer collisions occur, 
and the rescattering-generated flow is reduced, reducing
in magnitude the radial and elliptic flow and most 
of the HBT observables. Only the HBT parameter $R_{Long}$
increases for larger $\tau_{had}$ reflecting the
increased longitudinal size of the initial hadron source,
as seen in Equation 3. One can compensate for this
reduced flow in the other observables by introducing an
ad hoc initial ``flow velocity parameter'', but the increased 
$R_{Long}$ cannot be compensated by this new parameter.
In this sense, the initial hadron model used in the
present calculations with $\tau_{had}\sim1$ fm/c
and no initial flow is uniquely determined with
the help of $R_{Long}$.
\begin{figure}
\includegraphics{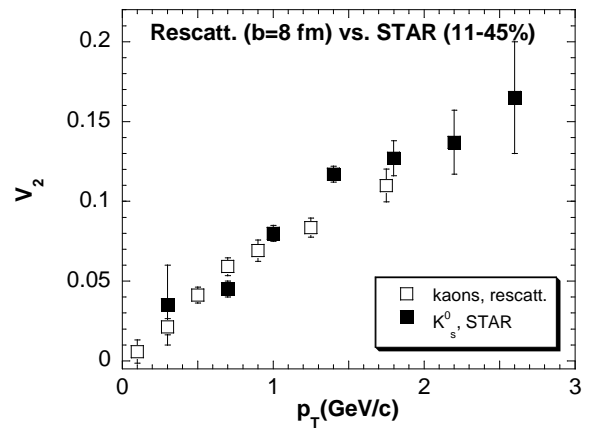}
\caption{\label{fig:kv2} Calculation of $v_2$ from the
rescattering model for kaons at $b=8$ fm compared with STAR
measurements for ${K^0}_s$ at 11-45\% centrality.}
\end{figure}

At this point, one can consider the physical significance of the
present rescattering model in two different ways. The first way is
to accept that it is physically valid in the time range where
hadronic rescattering should be valid, e.g. for times later
than when the particle density
reaches about 1 $fm^{-3}$, and to take the initial state hadronization
model as merely a parameterization useful to fit the data.
Considering the calculation this way, one can at least expect to
gain an insight into the phase-space configuration of the system
relatively early in the collision (in the calculation 1 $fm^{-3}$
occurs at a time 4 fm/c after hadronization). The second way
to consider the calculation is to see if it is possible to
also physically motivate the initial state hadronization model.
An attempt to do this is given below.

In order to consider the present initial state model as a physical 
picture, one must assume: 1) hadronization 
occurs very rapidly
after the nuclei have passed through each other, i.e.
$\tau_{had}=1$ fm/c, 2) hadrons or at least hadron-like 
objects can exist
in the early stage of the collision where the maximum value of
$\rho$ approaches 8 GeV/$fm^3$, and 3) the initial kinetic energies
of hadrons can be large enough to be described by $T=300$ MeV
in Equation 1.

Addressing assumption 2) first, in the calculation the maximum 
number density
of hadrons at mid-rapidity at $t=0$ fm/c is 6.8 $fm^{-3}$, 
rapidly dropping to about
1 $fm^{-3}$ at $t=4$ fm/c. Since most of these hadrons are pions,
it is useful as a comparison to estimate the effective volume of a
pion in the context of the $\pi-\pi$ scattering cross section,
which is about 0.8 $fm^2$ for s-waves \cite{Prakash:1993a}.
The ``radius'' of a pion is found to be 0.25 fm and
the effective pion volume is 0.065 $fm^3$, the reciprocal
of which is about 15 $fm^{-3}$. From this it is seen that at the 
maximum hadron number
density in the calculation, the particle occupancy 
of space is estimated to be
less than $50\%$, falling rapidly with time. One could
speculate that this may
be enough spacial separation to allow individual hadrons or
hadron-like objects to keep 
their identities and not melt into quark matter,
resulting in a ``super-heated'' semi-classical gas
of hadrons at very early times, as assumed in the
present calculation.
\begin{figure*}
\includegraphics{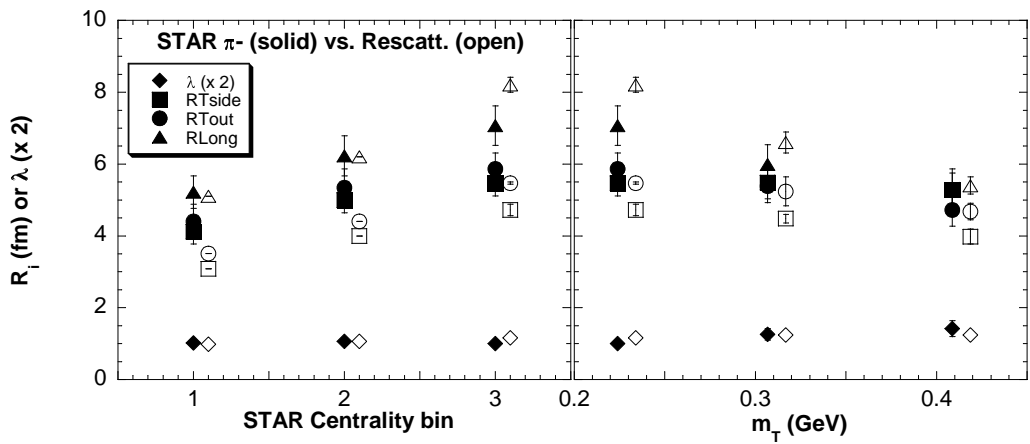}
\caption{\label{fig:hbt} Comparison of HBT source parameters from
rescattering with STAR measurements as a function of centrality
bin (see text) and $m_{T}$. The STAR measurements are the solid
symbols and the rescattering calculations are the open symbols.
The errors on the STAR measurements are statistical+systematic.}
\end{figure*}

Since the calculation takes the point of view of being purely hadronic,
it is instructive to consider assumptions 1) and 3) in the context
of the Hagedorn thermodynamic model of hadronic 
collisions \cite{Hagedorn:1970a}.
According to Hagedorn, the mass spectrum of hadrons of mass m
increases proportional to $\exp{(m/T_0)}$ in hadronic collisions, 
where $T_0=160$ MeV is the
limiting temperature of the system. This seems to
contradict the value $T=300$ MeV needed in the present case
in Equation 1 to describe the data.
The Hagedorn model assumes that a) the system comes to equilibrium
and b) the details of particle production via direct processes
and through resonance decay average out. Neither of these
assumptions is necessarily guaranteed at very early times in the
collision. The use of the thermal functional form, Equation 1, to
set up the initial transverse momenta of the hadrons in the present
calculations is convenient but not required. For example, 
the exponential 
form $\exp{(-m_T/T_e)}$ (where $T_e$ is a slope parameter) 
which does not describe thermal equilibrium,
could have equally well been used. This exponential form of the
transverse mass distribution was successfully used previously 
in rescattering
calculations to describe SPS data \cite{Humanic:1998a}.

Assumption 1) can also be motivated by the Color Glass
Condensate model \cite{McLerran:2002a,Kovner:1995a}.
In the usual version of this
picture, after the collision
takes place the Color Glass melts into quarks and gluons in
a timescale of about 0.3 fm/c at RHIC energy, and then the
matter expands and thermalizes into quark matter by about 
1 fm/c. In the context of the present rescattering calculations,
it is tempting to modify the collision senario such that 
instead of the
Color Glass melting into quarks and gluons just after
the collision, the sudden impact of the collision
``shatters'' it directly into hadronic
fragments on the same timescale as in the parton senario due 
to the hadronic strong interactions.

\section{Summary}

A simple thermal-like initial state model coupled
with a hadronic rescattering calculation
is able to adequately describe the large elliptic flow and 
small HBT radii recently measured at RHIC. A feature of this
picture is a very early hadronization time of about 1 fm/c
after the initial collision of the nuclei.

\begin{acknowledgments}
It is a pleasure to acknowledge Ulrich Heinz, Mike Lisa
and Rainer Renfordt for their helpful
suggestions regarding this work, and Larry McLerran 
for illuminating discussions on the Color Glass 
Condensate model. This work was supported by the U.S.
National Science Foundation under grant PHY-0099476.
\end{acknowledgments}


%


\end{document}